\documentclass[useAMS,usenatbib]{mn2e}

\usepackage{graphicx}	
\usepackage{amsmath}	
\usepackage{amssymb}	
\usepackage{float}
\usepackage{subfigure}
\usepackage{lscape}

\title[Formation of cE galaxies]{The formation of compact dwarf
  ellipticals through merging star clusters}   

\author[F. Urrutia Zapata et al.]{F. Urrutia Zapata$^{1}$\thanks{E-mail:
    feurrutia@udec.cl}, 
  M. Fellhauer$^{1}$,
  A.G. Alarc\'on Jara$^{1}$,
  D.R. Matus Carrillo$^{1}$,
  \newauthor and C.A. Aravena$^{1}$\\
  $^{1}$Departamento de Astronom\'ia, Universidad de Concepcion, Casilla
  160-C, 4030000 Concepci\'on, Chile}

\begin{document}

\label{firstpage}

\pagerange{\pageref{firstpage}--\pageref{lastpage}}
\pubyear{2018}

\maketitle

\begin{abstract}
In the last decades, extended old stellar clusters have been observed.   
These extended objects cover a large range in masses, from extended
clusters or faint fuzzies to ultra compact dwarf galaxies.  It has
been demonstrated that these extended objects can be the result of the
merging of star clusters in cluster complexes (small regions in which
dozens to hundreds of star clusters form). This formation channel is
called the ``Merging Star Cluster Scenario''.  This work tries 
to explain the formation of compact ellipticals in the same
theoretical framework.  Compact ellipticals are a comparatively rare
class of spheroidal galaxies, possessing very small effective radii
and high central surface brightnesses.  With the use of numerical
simulations we show that the merging star cluster scenario, adopted
for higher masses, as found with those galaxies, can reproduce all
major characteristics and the dynamics of these objects.  This opens
up a new formation channel to explain the existence of compact
elliptical galaxies.   
\end{abstract}


\section{Introduction}
\label{sec:intro}

In the last decades, extended old stellar clusters have been observed.
They are similar to globular clusters (GCs) but with larger sizes.  As
there is no clear division between the two classes of objects but
rather a smooth, continuous distribution of sizes, an arbitrary limit
of $R_{\rm eff} = 10$~pc is currently used to distinguish between the
different objects \citep[see e.g.][]{Bruens2012}.  Extended objects
(EOs) cover a wide range of masses.  Objects at the low mass end with
masses comparable to normal GCs are called extended clusters (ECs)
\citep[e.g.][]{Huxor2005, Chandar2004} or
faint fuzzies (FFs) \citep{Brodie2002, Burkert2005} and objects
at the high-mass end are called ultra compact dwarf galaxies
\citep[UCDs][]{Hilker1999,Drinkwater2000}.  Ultra compact dwarf
galaxies are compact objects with luminosities above the brightest known
GCs.  Again there is no clear boundary between high-mass GCs and
low-mass UCDs, so  usually, a lower mass limit of $2 \cdot
10^6$~M$_{\odot}$ is applied. 

A few decades ago young massive star clusters (YMCs) were found with GC
properties in gas-rich galaxies.  They are abundant in star-burst and
interacting galaxies, but they are also present in some apparently
unperturbed disk galaxies \citep{Larsen1999}.  Observations have shown
that YMCs are often not isolated, but are part of larger structures
\citep[e.g.][]{Whitmore1999} which were later dubbed cluster complexes
\citep{Bruens2009,Bruens2011}. The CCs contain from a few dozens to
hundreds of young massive star clusters spanning up to a few hundred
parsec in diameter.  The mass of a CC is the sum of the YMC in it.
Observations show that most CCs have a massive concentration of star
clusters in their centres and from a few to hundreds of isolated SC in
their vicinity.

\citet{Fellhauer2002a, Fellhauer2002b} demonstrated that objects like
ECs, FFs and UCDs can be the remnants of the merging of star cluster
complexes (CCs); they called it the Merging Star Cluster Scenario.  The
star clusters inside of CCs are bound to each other and therefore
experience constant encounters, which finally leads to a merging
process of most or even all SCs inside the complex.  The resulting
object is massive and has a larger effective radius than the single
constituents.  A more concise study was performed by
\citet{Bruens2009,Bruens2011}. 

Our work tries to explain the formation of compact ellipticals (cEs).
These objects are a comparatively rare class of spheroidal galaxies, 
possessing very small effective radii ($R_{\rm eff}$) defined as the
projected radius that encloses half of the total luminosity of 
one object.  The central surface brightness is high, compared to dwarf
ellipticals of the same size \citep{Faber1973}.  One of the most  
important characteristic of cEs is the high stellar density due the
high surface brightness and the small $R_{\rm eff}$.  The prototype
galaxy for this type of galaxy is the 32nd object in the catalog of
Messier \citep{Messier1784}, called M32, which is a satellite  
of the Andromeda galaxy (M31).  Until a few years ago only about a
dozen of these objects were identified up to a distance of about
$100$~Mpc but in the last years their number increased significantly
\citep{Chilingarian2015, Janz2016, Ferre2017}.

M32 has an $R_{\rm eff}$ of the bulge of $27$~arcseconds ($\sim
100$~pc), an effective surface brightness in the bulge of $18.23$
R-mag\,arcsec$^{-2}$ \citep{Graham2002} and a velocity dispersion of
$76 \pm 10$~km\,s$^{-1}$ \citep{Marel1998}. 

The standard formation scenario of these systems proposes a galaxy
origin.  CEs are the result of tidal stripping and the truncation of
nucleated larger systems \citep{Faber1973}.  The most recent
  simulations by \citet{Du2018} show that by including ram pressure
  into the simulations the high metallicity of cEs can be explained.
Or they could be a natural extension of the class of elliptical
galaxies to lower luminosities and smaller sizes  \citep{Wirth1984,
  Kormendy2009}.  \citet{Kormendy2012} argue against the stripping
scenario as ``not all cEs are companions of massive galaxies''.

\citet{Huxor2011} report two cEs which show evidence of formation
resulting from ongoing tidal stripping of more massive progenitors,
i.e.\ the tidal tails are clearly visible.  But later
\citet{Huxor2013} found the first isolated cE.  Its isolated position
suggests that the stripping scenario may not be the only possible
formation channel.  \citet{Chilingarian2015} report that isolated
  cEs may be ejected from their host systems.

We want to propose a completely new formation scenario for cEs.  In
our project we try to model cEs in a similar way like UCDs in
\citet{Fellhauer2002a}, which are objects with characteristics very
similar to cE: high surface brightness and small $R_{\rm eff}$, i.e.\
high stellar density, using the merging star cluster scenario extended
to higher masses and sizes.  We think that in the early Universe we
might have produced sufficiently strong star bursts to form cluster
complexes which merge into cEs.  So far it is observationally unknown
if cEs are dark matter dominated objects.  If our scenario is true, 
then they (or at least some of them) would be dark matter free, very
extended and massive "star clusters". 

\section{Setup}
\label{sec:setup}

For our cE models we divide the total mass of the cluster complex of
$M_{\rm cc} = 10^9$~M$_\odot$ (this initial mass of the CC is kept constant 
in all our simulations) into $N_{0} = 64$ or $128$ 'UCDs':  
\begin{eqnarray}
  \label{eq:1a}
  M_{\rm sc,64} & = & \frac{M_{\rm cc}}{N_{0}} \ = \ \frac{10^9 {\rm
      M_{\odot}}}{64} \ = \ 15,625,000 \ {\rm M_{\odot}}, \\
  \label{eq:1b}
  M_{\rm sc,128} & = & \frac{10^9 {\rm M_{\odot}}}{128} \ = \ 7,812,500
  \ {\rm M_{\odot}}.
\end{eqnarray}
As one can see, the term UCD is justified because our star clusters
inside the CC have masses larger than the adopted mass-limit for
UCDs.  As this and previous studies have shown, the number of objects
$N_{0}$ does not alter the results and therefore we are confident to
obtain qualitatively similar results if we would have divided the
total mass of the CC into more and smaller constituents.  The
  same reasoning applies for all SCs having the same mass, even though
observations show a power-law mass-spectrum $\propto M^{-2}$.
Applying the correct mass function would mean that for every massive
SC of mass $10^{6}$~M$_{\odot}$ we would have to simulate 100 SCs of
$10^{5}$ and even 10,000 SCs of $10^{4}$~M$_{\odot}$.  This is beyond
the capacities of the used computer systems.  Smaller simulations
using a shallower mass spectrum are reported in \citet{Fellhauer2003}
and show no differences.

The constituents (star clusters or UCDs) are modeled as Plummer
spheres \citep{Plummer1911} using $100,000$ particles.  The size of
the SCs $R_{\rm sc}$ is varied to be $4$, $10$ or $20$~pc, with masses
according to Eqs.~\ref{eq:1a} and \ref{eq:1b}.  These scale radii
cover the observed values of YMC and most UCDs
\citep{Whitmore1999,Maraston2004}. The star clusters then
are distributed inside the CC according to another Plummer
distribution with a scale radius of $R_{\rm cc}$ being $50$, $100$ or
$200$~pc.  This choice reflects sizes seen with CCs in various
  observations \citep{Whitmore1999, Bastian2005}.  It also is the range
of initial values which could produce final merger objects similar to
cEs.  For all Plummer distributions we choose a cut-off radius equal
to five Plummer radii.  An overview over this parameters is given in
Tab.~\ref{tab:para}. 

\begin{table}
  \centering
  \caption{Parameters varied in the simulations.  The distance to the
    center of the galaxy of the CC ($R_{gal}$),  the Plummer radius of
    the SCs ($R_{sc}$), the Plummer radius of the cluster complex
    ($R_{cc}$) and the initial number of 'UCDs' in the CC ($N_{0}$).} 
  \label{tab:para}
  \begin{tabular}{ccrrr}
    \hline
    $R_{\rm gal}$ & [kpc]& 20 & 60 & 100\\
    $R_{\rm sc}$  & [pc] & 4  & 10 & 20\\
    $R_{\rm cc}$  &[pc]  & 50 & 100 & 200 \\
    $N_{0}$      &      & 64 & 128 &\\
    \hline
  \end{tabular}
\end{table}

We simulate models with and without tidal field (at different galactic
distances).  The tidal field is modeled as a standard galactic
potential as it is often used to model the Milky Way
\citep{Mizutani2003} leading to a rotational velocity of about
$220$~km\,s$^{-1}$ at the solar radius and an almost flat rotation
curve further out.  The chosen galactic distances and the resulting
parameters are shown in Tab.~\ref{tab:galpot}. 

\begin{table}
  \centering
  \caption{Parameters for the chosen orbits.  First column shows the
    distance to the centre of the galaxy of the CC.  Second column
    shows the circular velocity at this distance.  Third column gives
    the time the CC and later the merger object needs to perform one
    orbit around the galaxy.  Last column gives the number of orbits
    the simulation performs, given a standard simulation time of
    $10$~Gyr.}
  \label{tab:galpot}
  \begin{tabular}{rrrr}
    \hline
    $R_{\rm gal}$ & $v_{\rm circ}$ & $T_{\rm orb}$ & $N_{\rm revol.}$ \\
    $[$kpc] & [km\,s$^{-1}$] & [Gyr] & \\ \hline
     20 & 230.0 & 0.534 & 18.7 \\
     60 & 206.6 & 1.778 &  5.6 \\
    100 & 219.8 & 2.795 &  3.6 \\
    \hline
  \end{tabular}
\end{table}

There $R_{\rm gal}$ is the distance to the center of the galaxy of the
distribution of SCs/UCDs, $v_{\rm circ}$ is the initial velocity of
the distribution to have a circular orbit, T$_{\rm orb}$ is the
orbital period and $N_{\rm revol}$ is the number of revolutions and
it is given by $\frac{\rm t_{\rm int} = 10 {\rm Gyr}}{\rm T_{\rm orb}}$, 
where $t_{\rm int}$ is our simulations time which is always $10$~Gyr.

We perform the simulations using the particle-mesh code {\sc Superbox}
\citep{Fellhauer2000}.  {\sc Superbox} is a particle-multi-mesh code 
with high-resolution sub-grids, which stay focused on the moving 
objects.

To get a good resolution of the star clusters, the individual high
resolution grids cover an entire star cluster, whereas the medium
resolution grids of every star cluster embed the whole initial CC.  
The outermost grid covers the whole orbit of the CC around the
galaxy. 

We perform in total 162 simulations.  81 simulations with initially
$N_0 = 64$ and then similar simulations with $N_0 = 128$.  For each
combination of $R_{\rm gal}$, $R_{\rm sc}$ and $R_{\rm cc}$ (27
possible combinations) we perform three random realisations of the
positions and velocities of the SCs inside the CC to assess the
possible spread of our results. 

\section{Results}
\label{sec:results}

\begin{figure}
  \centering
  \includegraphics[width=9.0cm]{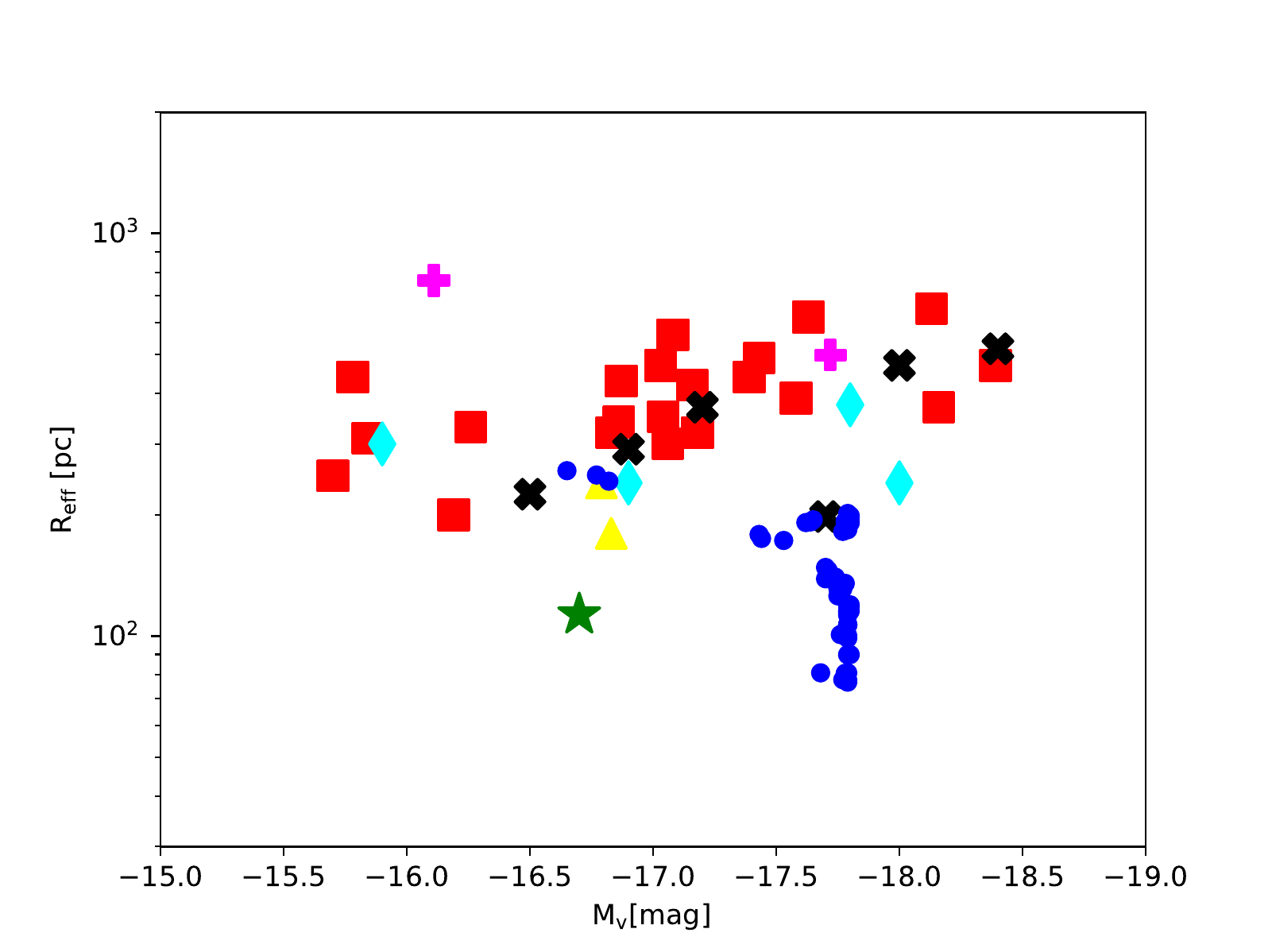}
  \caption{Absolute Magnitude vs.\ effective radius of observed cEs
     and our simulations.  The green star denotes M32.  The symbols are
     the different observational values taken from
     \citet{Chilingarian2007} (yellow triangles), \citet{Chilingarian2009}
     (red squares), \citet{Huxor2013} (magenta plus), \citet{Misgeld2011} 
     (black crosses) and \citet{Norris2014} (cyan diamonds).  
     The blue circles are our simulation results.  Note that in order to 
     determine the absolute magnitude we applied a generic M/L-ratio of 
     unity, i.e.\ our objects should be less bright in reality.  Furthermore, 
     all our models have the same initial mass, therefore the final mass (or
     brightness) is also almost the same.}
  \label{fig:obs}
\end{figure}

In all simulations, the merging process leads to a stable object which
we are analysing after $10$~Gyr of simulation time.

\begin{table*}
  \centering
  \caption{Results of the simulations of cEs.  The name of the
    simulations (first column) is according to the size of a single
    SC/UCD ($R_{\rm sc}$; $4$=A, $10$=B, $20$=C), the size of the
    distribution or CC ($R_{\rm cc}$; $50$=a, $100$=b, $200$=c) and
    the distance to the centre of the galaxy of the distribution
    ($R_{\rm gal}$; $20$=1, $60$=2, $100$=3).  Columns 2, 3 and 4 give
    the varied setup parameters.  Columns 5--8 give the number of SCs
    ending up in the merger object ($N_{\rm merger}$) and the final mass
    of the merger object ($M_{\rm merger}$) both for $N_{0} = 64$ and
    $128$ SCs in the distribution.  Column 9 gives the effective
    radius, column 10 the ellipticity, column 11 the central velocity
    dispersion and column 12 the central surface brightness of the
    merger object for $N_{0} = 64$; columns 13--16 the same values for
    $N_{0} = 128$.}
  \label{tab:results}
  \begin{tabular}{lrrrrrrrrrr} \hline
    Name & $R_{\rm sc}$ & $R_{\rm cc}$ & $R_{\rm gal}$ & $N_{0}$ &
    $N_{\rm merger}$ & $M_{\rm merger}$ & $R_{\rm eff}$ & $\epsilon$
    & $\sigma_{0}$ & $\Sigma_{0}$ \\
    & [pc] & [pc] & [kpc] & & & [$10^{7}$~M$_{\odot}$] & [pc] & &
    [km\,s$^{-1}$] & [mag\,arcsec$^{-2}$] \\ \hline
    SimAa1 & 4 & 50 & 20 & 64 & 61.6$\pm$0.5 & 96$\pm$1.3 &
    131.4$\pm$13.4 & 0.1$\pm$0.09 & 56.6$\pm$0.8 & 15.6$\pm$0.1 \\
    SimAa2 & 4 & 50 & 60 & 64 & 62.3$\pm$1.5 & 95.1$\pm$3.9 &
    140$\pm$10 & 0.05$\pm$0.04 & 56.2$\pm$0.8 & 15.8$\pm$0.2 \\
    SimAa3 & 4 & 50 & 100 & 64 & 62.6$\pm$1.5 & 91.6$\pm$1.5 &
    138.7$\pm$10.3 & 0.08$\pm$0.07 & 55.5$\pm$0.4 & 15.8$\pm$0.2 \\
    SimBa1 & 10 & 50 & 20 & 64 & 63.6$\pm$0.5 & 99.5$\pm$0.9 &
    98.7$\pm$11.3 &0.1$\pm$0.03 &72$\pm$4.3   &14.9$\pm$0.2 \\
    SimBa2 & 10 & 50 & 60 & 64 & 63.6$\pm$0.5 &99.1$\pm$1.1 &
    100.2$\pm$11.3 &0.09$\pm$0.02&71.4$\pm$3.3 & 14.9$\pm$0.2 \\
    SimBa3 & 10 & 50 & 100 & 64 & 63.6$\pm$0.5 & 96.9$\pm$1.9 &
    100.8$\pm$11.2 & 0.09$\pm$0.02 & 71.2$\pm$3.6 & 14.9$\pm$0.2 \\
    SimCa1 & 20 & 50 & 20 & 64 & 64 & 100 &
    89.9$\pm$4.4 & 0.09$\pm$0.07 & 82.3$\pm$4.1 & 14.9 \\
    SimCa2 & 20 & 50 & 60 & 64 & 64 & 99.6$\pm$0.5 & 
    89.8$\pm$4.4 & 0.14$\pm$0.01 & 82.6$\pm$4.1 & 14.9$\pm$0.1 \\
    SimCa3 & 20 & 50 & 100 & 64 & 64 & 99.1$\pm$0.6 &
    89.9$\pm$4.5 & 0.17$\pm$0.08 & 81.6$\pm$2.1 & 14.9 \\
    \hline
    SimAb1 & 4 & 100 & 20 & 64 & 50.3$\pm$11.5 & 78.5$\pm$17.9 &
    172.7$\pm$30.2 & 0.18$\pm$0.02 & 50.2$\pm$8.5 &16.1$\pm$0.4 \\
    SimAb2 & 4 & 100 & 64 & 60 & 47.3$\pm$14.5 & 72.1$\pm$20.7 &
    174.7$\pm$36 & 0.18$\pm$0.04 & 49.6$\pm$9.1 & 16.1$\pm$0.4 \\
    SimAb3 & 4 & 100 & 100 & 64 & 50$\pm$12 & 71.6$\pm$14.8 &
    178.7$\pm$35.4 & 0.21$\pm$0.16 & 49.2$\pm$9.3 &16.1$\pm$0.4 \\
    SimBb1 & 10 & 100 & 20 & 64 & 64 & 100 & 
    119.5$\pm$26.8 & 0.16$\pm$0.08 & 68.1$\pm$5.4 & 15$\pm$0.2 \\
    SimBb2 & 10 & 100 & 60 & 64 & 64 & 100 &
    118.3$\pm$25.3 & 0.17$\pm$0.12 & 67.5$\pm$4.4 & 15.1$\pm$0.2 \\
    SimBb3 & 10 & 100 & 100 & 64 & 64 & 99.4$\pm$0.6 &
    118.6$\pm$25.4 & 0.1$\pm$0.05 & 66.1$\pm$6.2 & 15.1$\pm$0.2 \\
    SimCb1 & 20 & 100 & 20 & 64 & 64 & 100 &
    115.3$\pm$8.3 & 0.19$\pm$0.18 & 71.7$\pm$5.4 & 15.2$\pm$0.1 \\
    SimCb2 & 20 & 100 & 60 & 64 & 64 & 99.9$\pm$0.1 &
    115.2$\pm$8.2 & 0.19$\pm$0.13 & 71$\pm$4.6 & 15.2$\pm$0.2 \\
    SimCb3 & 20 & 100 & 100 & 64 & 64 & 99.8$\pm$0.1 &
    115.2$\pm$8.3 & 0.09$\pm$0.09 & 71.1$\pm$4.8 & 15.2$\pm$0.2 \\
    \hline
    SimAc1 & 4 & 200 & 20 & 64 & 26$\pm$13 & 40.6$\pm$20.3 &
    242.4$\pm$30.4 & 0.2$\pm$0.09 & 27.1$\pm$8.6 & 17.3$\pm$0.2 \\
    SimAc2 & 4 & 200 & 60 & 64 & 26$\pm$13 & 38.9$\pm$18.9 & 
    251.1$\pm$43.2 & 0.18$\pm$0.08 & 26.5$\pm$7.3 & 17.2$\pm$0.2 \\
    SimAc3 & 4 & 200 & 100 & 64 & 25$\pm$12.1 & 34.8$\pm$16.4 &
    257.4$\pm$23.9 & 0.18$\pm$0.03 & 26.2$\pm$7.2 &17.2$\pm$0.2 \\
    SimBc1 & 10 & 200 & 20 & 64 & 64 & 100 &
    197.2$\pm$62.6 & 0.13$\pm$0.03 & 52.9$\pm$4.5 & 15.5$\pm$0.2 \\
    SimBc2 & 10 & 200 & 60 & 64 & 64 & 100$\pm$0.1 &
    199$\pm$67.8 & 0.14$\pm$0.03 & 51.9$\pm$5.9 & 15.5$\pm$0.1 \\
    SimBc3 & 10 & 200 & 100 & 64 & 64 & 99.5$\pm$0.1 &
    201.5$\pm$65.5 & 0.14$\pm$0.02 & 51.4$\pm$4.6 & 15.4$\pm$0.1 \\
    SimCc1 & 20 & 200 & 20 & 64 & 64 & 100 &
    192.7$\pm$49.3 & 0.32$\pm$0.11 & 55.1$\pm$6 & 15.9$\pm$0.3 \\
    SimCc2 & 20 & 200 & 60 & 64 & 64 &100 &
    190.8$\pm$44.9 & 0.24$\pm$0.13 & 54.4$\pm$5.7 & 15.9$\pm$0.2 \\
    SimCc3 & 20 & 200 & 100 & 64 & 64 & 99.8$\pm$0.1 &
    190.5 $\pm$42.4 & 0.21$\pm$0.14 & 53.9$\pm$5.5 & 15.9$\pm$0.3 \\
    \hline
    \hline
    SimAa1 & 4 & 50 & 20 & 128 & 125.7$\pm$ 0.6 & 98.2$\pm$0.5 &
    135.1$\pm$9.1 & 0.07$\pm$0.05 & 58.2$\pm$2.3 & 15.4$\pm$0.1 \\
    SimAa2 & 4 & 50 & 60 & 128 & 125.7$\pm$0.6 & 98  $\pm$0.5 & 
    131$\pm$7.3 & 0.07$\pm$0.05 & 58.7$\pm$2.8 & 15.5$\pm$0.1 \\
    SimAa3 & 4 & 50 & 100 & 128 & 125.7$\pm$ 0.6 & 96.3$\pm$1.6 &
    125.8$\pm$12.5 & 0.09$\pm$0.05 & 58.6$\pm$2.6 & 15.5$\pm$0.1\\
    SimBa1 & 10 & 50 & 20 & 128 & 127 & 99.2 &
    76.8$\pm$11.2  & 0.1 $\pm$0.04  & 72  $\pm$4.6 & 14.8$\pm$0.1 \\ 
    SimBa2 & 10 & 50 & 60 & 128 & 127.3$\pm$ 0.6 & 99.4$\pm$0.5 &
    77.7$\pm$11 & 0.16$\pm$0.1 & 73.8$\pm$4.3 & 14.8$\pm$0.2 \\ 
    SimBa3 & 10 & 50 & 100 & 128 & 127.3$\pm$ 0.6 & 98 $\pm$1.7 &
    77.9$\pm$11 & 0.19$\pm$0.1 & 74.1$\pm$6.1 & 14.8$\pm$0.2 \\ 
    SimCa1 & 20 & 50 & 20 &128 & 127.7$\pm$ 0.6 & 99.7$\pm$0.5 &
    81$\pm$10.2 & 0.1$\pm$0.1 & 85.5$\pm$0.4 & 14.8$\pm$0.1 \\   
    SimCa2 &20&50 &60 & 128 & 99.4$\pm$0.4 & 89.8$\pm$4.4 &
    81$\pm$10.2 & 0.11$\pm$0.07 & 85.2$\pm$0.9 & 14.8$\pm$0.1 \\
    SimCa3 & 20 & 50 & 100 & 128 & 127.3$\pm$ 0.6 & 98.9$\pm$0.3 &
    80.8$\pm$9.9 & 0.08$\pm$0.09 & 83.6$\pm$1.3 &14.9$\pm$0.1 \\
    \hline
    SimAb1 & 4 & 100 & 20 & 128 & 118$\pm$14 & 92.2$\pm$10.9 &
    144.2$\pm$12.1 & 0.09$\pm$0.1 & 58.8$\pm$1.3 & 15.7$\pm$0.2 \\ 
    SimAb2 & 4 & 100 & 60 & 128 & 118.7$\pm$ 12.7 & 92.7$\pm$10 &
    145.7$\pm$12.2 & 0.1$\pm$0.02 & 59.1$\pm$1.3 & 15.7$\pm$0.1 \\ 
    SimAb3 & 4 & 100 & 100 & 128 & 118.7$\pm$ 13.7 & 91.9$\pm$10.4 &
    148$\pm$12.6 & 0.13$\pm$0.04 & 57.8$\pm$0.1 &15.7$\pm$0.1 \\ 
    SimBb1 & 10 & 100 & 20 & 128 & 127.7$\pm$0.6 & 99.7$\pm$0.5 &
    106.6$\pm$20.6 & 0.11$\pm$ 0.08 & 69.3$\pm$2.9 & 15.1$\pm$0.1 \\ 
    SimBb2 & 10 & 100 & 60 & 128 & 127.7$\pm$ 0.6 & 99.7$\pm$0.5 &
    106.6$\pm$20.7 & 0.19$\pm$ 0.08 & 67.3$\pm$3.2 & 15$\pm$0.2 \\ 
    SimBb3 & 10 & 100 & 100 & 128 & 127.7$\pm$ 0.6 & 99.4$\pm$0.4 &
    106.6$\pm$20.6 & 0.17$\pm$0.06 & 67.8$\pm$2.9 & 15$\pm$0.1 \\
    SimCb1 & 20 & 100 & 20 & 128 & 127.3$\pm$0.6 & 99.5$\pm$0.5 &
    112.7$\pm$12.5 & 0.17$\pm$0.1 & 75.1$\pm$4.7 & 15.2$\pm$0.2\\ 
    SimCb2 & 20 & 100 & 60 & 128 & 127.7$\pm$0.6 & 99.7$\pm$0.4 &
    113$\pm$12.8 & 0.17$\pm$ 0.12 & 75$\pm$4.1 & 15.2$\pm$0.1 \\ 
    SimCb3 & 20 & 100 & 100 & 128 & 127.7$\pm$0.6 & 99.7$\pm$0.4 &
    114$\pm$14.3 & 0.21$\pm$0.18 & 76.5$\pm$3.4 & 15.3$\pm$0.2 \\ 
    \hline
    SimAc1 & 4 & 200 & 20 & 128 & 112.3$\pm$8.7 & 87.8$\pm$6.8 &
    194.3$\pm$16.7 & 0.08$\pm$0.02 & 43.8$\pm$6.8 & 16.1$\pm$0.2 \\
    SimAc2 & 4 & 200 & 60 & 128 & 110.7$\pm$10.6 & 86.4$\pm$8.3 &
    192.1$\pm$13.7 & 0.04 & 44.5$\pm$6.1 & 16.2$\pm$0.2 \\ 
    SimAc3 & 4 & 200 & 100 & 128 & 110.7$\pm$8.6 & 85.4$\pm$6.5 &
    191.3$\pm$16.7 & 0.09$\pm$0.03 & 43.1$\pm$6.3 & 16.1$\pm$0.2 \\
    SimBc1 & 10 & 200 & 20 & 128 & 126.7$\pm$1.5 & 99$\pm$1.2 &
    192.7$\pm$43.8 & 0.12$\pm$0.06 & 52.1$\pm$9.3 & 15.9$\pm$0.4 \\ 
    SimBc2 & 10 & 200 & 60 & 128 & 126$\pm$2 & 98.4$\pm$1.6 &
    192.3$\pm$39 & 0.13$\pm$0.09 & 52.5$\pm$8.3 & 15.8$\pm$0.2 \\
    SimBc3 & 10 & 200 & 100 & 128 & 127$\pm$1 & 99.1$\pm$0.7 &
    193.7$\pm$43 & 0.1$\pm$0.04 & 52$\pm$8.7 & 15.9$\pm$0.3 \\ 
    SimCc1 & 20 & 200 & 20 & 128 & 125$\pm$1 & 97.7$\pm$0.8 &
    181.9$\pm$46.6 & 0.19$\pm$0.05 & 54.8$\pm$5.5 & 16$\pm$0.2 \\ 
    SimCc2 & 20 & 200 & 60 & 128 & 127$\pm$1.7 & 99.2$\pm$1.4 &
    183.6$\pm$45.6 & 0.22$\pm$0.1 & 55.4$\pm$6.5 & 15.9$\pm$0.3 \\
    SimCc3 & 20 & 200 & 100 & 128 & 127$\pm$1.7 & 99.2$\pm$1.4 &
    184.9$\pm$45.3 & 0.21$\pm$0.11 & 55.5$\pm$6.3 & 15.9$\pm$0.2 \\
    \hline
  \end{tabular}
\end{table*}

\begin{figure*}
  \centering
  \includegraphics[width=8cm]{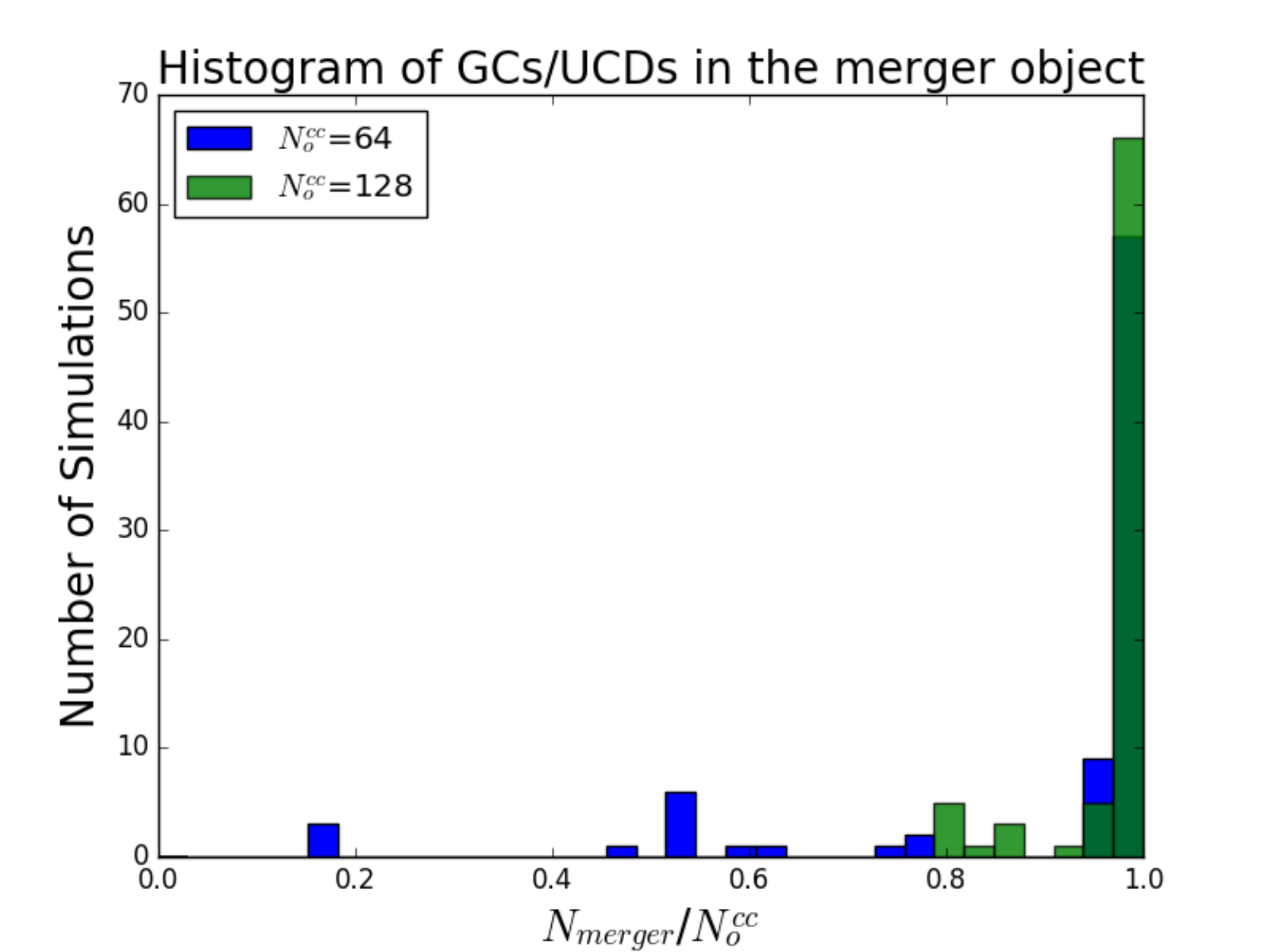}
  \includegraphics[width=8cm]{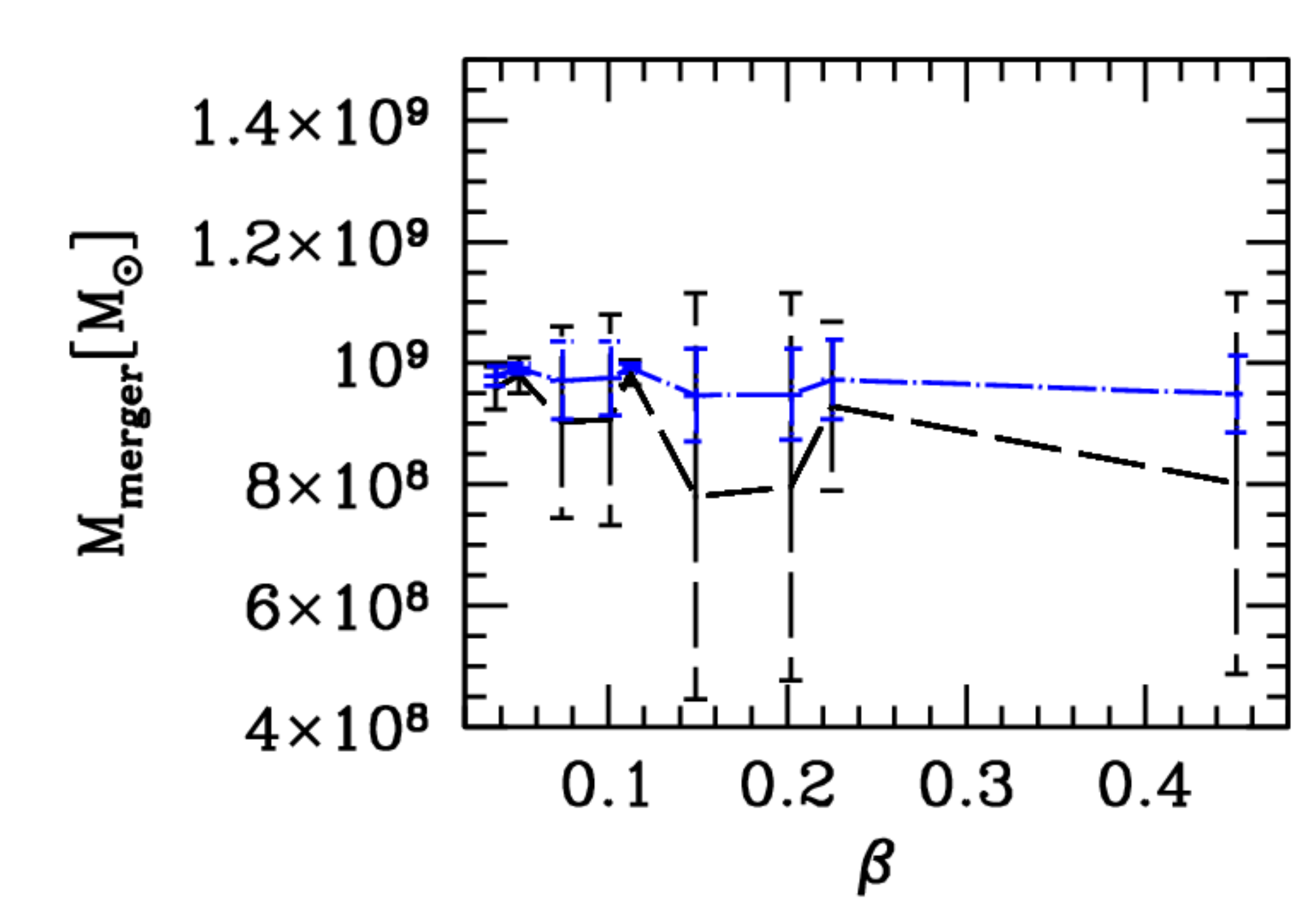}
  \caption{Left panel: Number of simulations vs. $N_{\rm merger}/N_{0}$
    (total number of simulations is 162).  Right Panel: Mass of the
    final object against the tidal filling parameter $\beta = 5 \cdot
    R_{\rm cc} / r_{\rm tidal}$.  It is obvious that our parameter
    space produces tidally under-filling initial conditions and
    therefore almost all the mass of the CC ends up in the final
    object.} 
  \label{fig:Nmerger}
\end{figure*}

\begin{figure*}
  \centering
  \includegraphics[width=8cm]{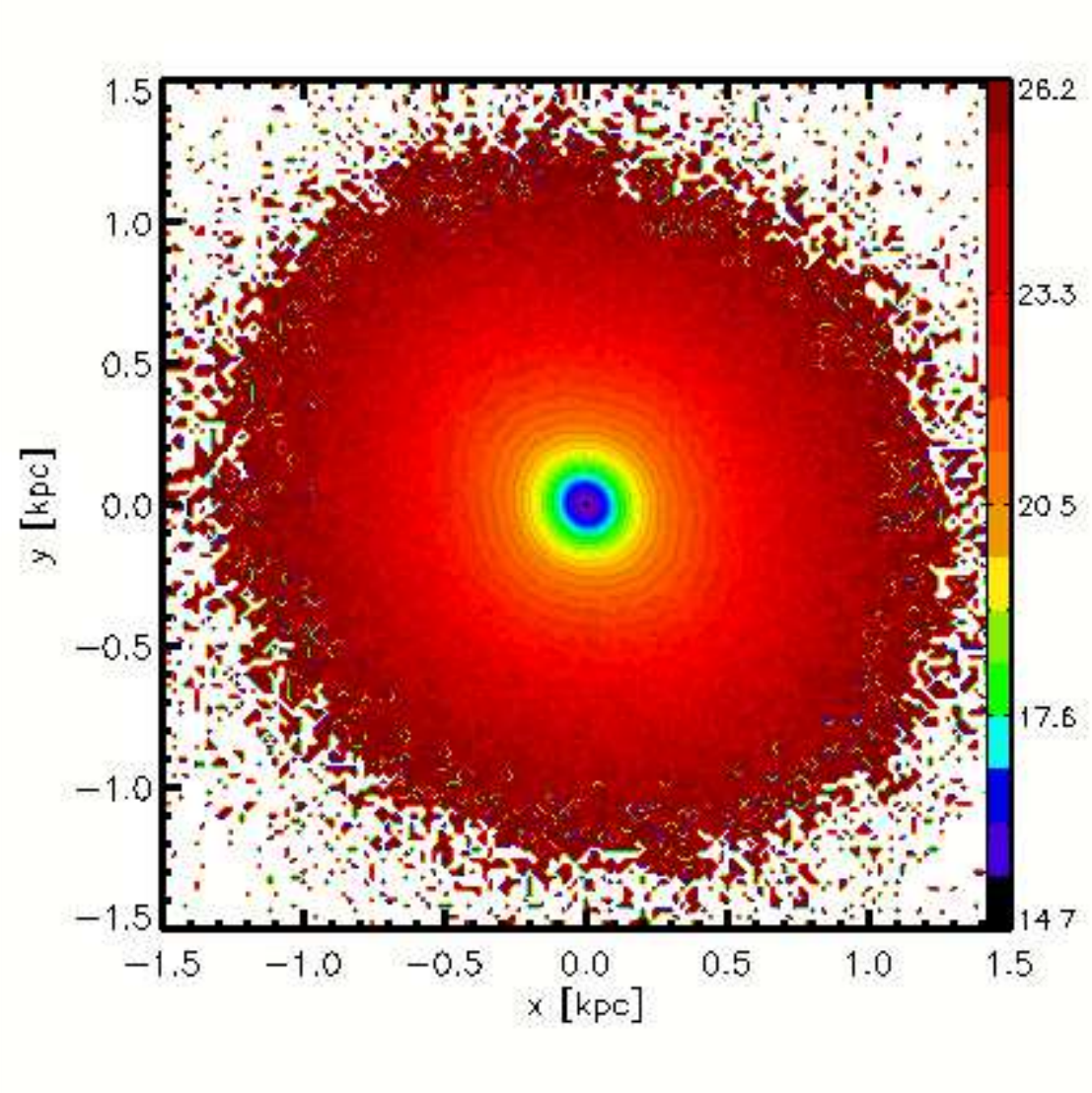}
  \includegraphics[width=8cm]{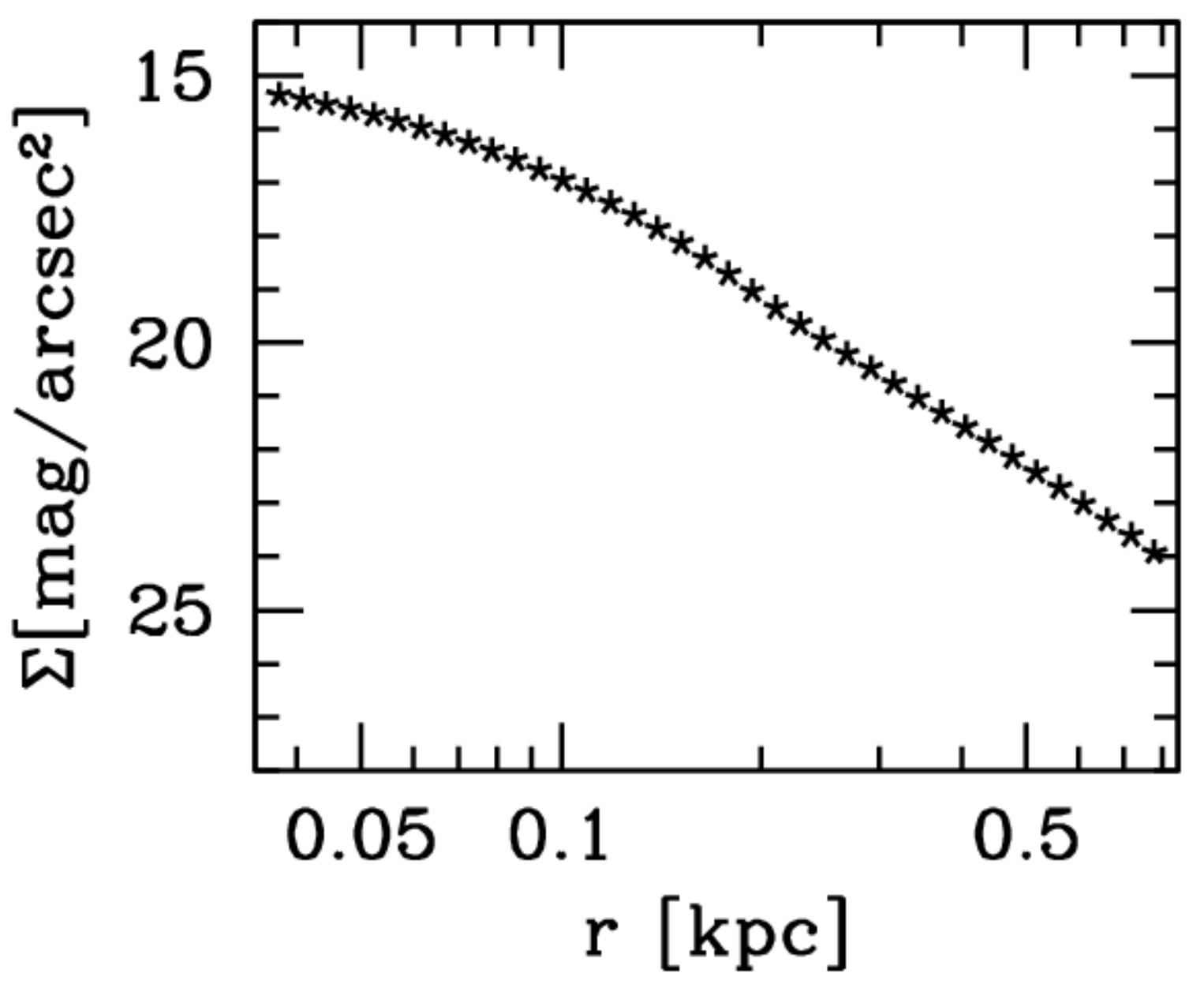}
  \caption{Sim85: $R_{\rm sc} = 10$~pc, $R_{\rm cc} = 50$~pc.  Left
    panel: 2D pixel-map of surface brightness for one of our objects.
    The magnitudes are derived using a M/L-ratio of unity.  Right
    panel: Radial profile of surface brightness calculated using
    concentric, logarithmically spaced rings.}
  \label{fig:SimulationExample}
\end{figure*}

We determine the number of constituents, which merge to build
the final object $N_{\rm merger}$ and measure the final bound mass of
this object $M_{\rm merger}$.

To obtain the effective radius $R_{\rm eff}$ of our object we fit a
Sersic profile \citep{Sersic1963} of the form
\begin{eqnarray}
  \label{eq:sersic}
  I(R) & = & I_{\rm e} \exp \left \{ - b_{n} \left[ \left(
        \frac{R}{R_{\rm eff}} \right)^{1/n} - 1 \right] \right\},
\end{eqnarray}
with $b_n$ satisfying the next equation which relates the incomplete
Gamma function ($\gamma$) and the Gamma function ($\Gamma$) 
\begin{eqnarray}
\label{eq:bn}
 \gamma(2n;b_{n}) & = & \frac{1}{2}\Gamma(2n).
\end{eqnarray}

The central surface brightness we determine by producing a 2D
pixel-map of the surface densities of our object, searching for the
pixel with the highest value and converting this value into magnitudes
per arcseconds$^{2}$ by using a generic mass-to-light (M/L) ratio of
unity.  As we have no information in our simulations about age
  and metallicity of our stars (our particles are indeed rather
  phase-space representations), we refrain from applying any more
  realistic values for an old population.

The ellipticity of our object is taken from the routine {\sc Ellipse}
in IRAF.  As input we use the 2D pixel map we already produced from
our simulation results.  We use the ellipticity of the nearest
isophote to a generic distance of $100$~pc.

Finally, we use all particles in the central pixel to determine the
central velocity dispersion of our object.

An overview of the combinations of setup parameters and the mean
values of our results can be found in Tab.~\ref{tab:results}.

As Tab.~\ref{tab:results} shows, we do not see any significant
difference between the results obtained if we distribute the mass of
the CC into $64$ objects or $128$.  With this we do agree with
previous studies \citep[e.g.][]{Bruens2009}.  Therefore, in order to
boost our statistics we use in the following analysis of the results
mean values obtained by adding the two sets of simulations together.

Furthermore, we see that almost all sets of parameters used in our
study lead to objects similar to cEs.  In Fig.~\ref{fig:obs} we
compare our results with data from various observations in plotting
total brightness against effective radius.  Our simulations have
effective radii similar to M32 and total masses similar to the mean
value of compact ellipticals.  Please note that we have used a generic
M/L ratio of unity and therefore we overestimate the total luminosity
of our objects.  Furthermore, as we keep the mass of our CC constant
and almost all SC/UCDs end up in the merger object, almost all of our
results show approximately the same absolute magnitude.

\subsection{SCs/UCDs in the merger object}
\label{sec:Nmerger}

In the left panel of Fig.~\ref{fig:Nmerger} we show in how many
simulations we obtain a certain fraction of star clusters/UCDs merging 
and ending up in the final object as a histogram.  The blue bars
represent simulations with $64$ constituents and the green bars
$128$.  It is clearly visible that in almost all simulations the
majority if not all objects merge and build the final compact
elliptical.  

This behaviour can be understood well by looking at the right panel of
Fig.~\ref{fig:Nmerger}.  Here we show the final mass of the merger
object as function of the tidal filling parameter $\beta$. 
For $\beta$ we use the definition from \citet{Fellhauer2002}:
\begin{eqnarray}
  \label{eq:beta}
  \beta & = & \frac{R_{\rm cut,cc}}{r_{\rm tidal}}
\end{eqnarray}
where $R_{\rm cut,cc}$ is the cut-off radius of the SC/UCD
distribution, which is chosen to be $R_{\rm cut,cc} = 5 \cdot R_{\rm
  cc}$, and $r_{\rm tidal}$ is the tidal radius of this distribution
at the given distance from the galaxy and for the used potential.  The
tidal radius is calculated using the standard equation from
\citep{Binney1987}: 
\begin{eqnarray}
  \label{eq:tidal}
  r_{\rm tidal} & = & \left( \frac{M_{\rm cc} = 10^{9}{\rm M}_{\odot}}{3 M_{\rm
        gal}(R_{\rm gal})} \right)^{1/3} R_{\rm gal}.
\end{eqnarray}

In our simulations the $\beta$ values are all below $0.5$.  This means
that all our initial models are extremely tidally under-filling.  From
this fact, it is clear that most constituents should end up in the
final merger object.  The old study by \citet[][, their figure
8]{Fellhauer2002} even showed that $\beta$ values slightly above unity
still lead to almost all SCs merging.

In their figure~8 it is also shown that one might expect to see
differences in this merging behaviour according to the filling factor
parameter $\alpha$, which the authors define as:
\begin{eqnarray}
  \label{eq:alpha}
  \alpha & = & \frac{R_{\rm sc}}{R_{\rm cc}}.
\end{eqnarray}
The less filled the CC is with star clusters the lower should be the
number of merged star clusters.  We see this secondary trend in our
simulations as well.  The few simulations which have low $N_{\rm
  merger}$ values have also very low values of $\alpha$, i.e.\ these
are mainly the simulations with $R_{\rm sc} = 4$~pc and $R_{\rm cc} = 100
$ or $200$~pc. 

According to \citet{Fellhauer2002} at low $\alpha$ values the small
constituents are merging first with each other and only later build up
the central merger object, while at higher values of $\alpha$ the
merging happens fast and mainly with the central object.

Another mechanism to lose SCs from the CC is due to the random
velocities they have.  It is therefore possible that a SC is placed in
the outskirts of the CC distribution with an outwards velocity, i.e.\
an apo-centre of its orbit beyond $R_{\rm cut,cc}$ and may have the
possibility to escape.

With our choice of the initial mass of the CC, it is clear that the
simulations with very incomplete merging processes, i.e.\ many
escaping star clusters can not resemble a final object with properties
of a compact elliptical, as the final mass is far below of what we expect
of a cE.

\subsection{Shapes}
\label{sec:shapes}

In Fig.~\ref{fig:SimulationExample} we show in the left panel a 2D
smoothed pixel-map of one of our objects, which resembles a cE galaxy.
We have converted the mass per area values of surface density from our
simulation into magnitudes per solid angle using a generic
mass-to-light ratio of unity.  The figure shows clearly that the
product of our simulation setup is a massive and compact object with a
size and luminosity similar to what has been observed in compact
elliptical galaxies.

In the right panel we show the radial surface brightness profile.  The
values are obtained by analysing the surface density of our merger
object in concentric rings, which are logarithmically spaced.  As most 
of our objects exhibit low ellipticities we are confident that
the error by using rings instead of ellipses is of no significance.

\begin{figure}
  \centering
  \includegraphics[width=8cm]{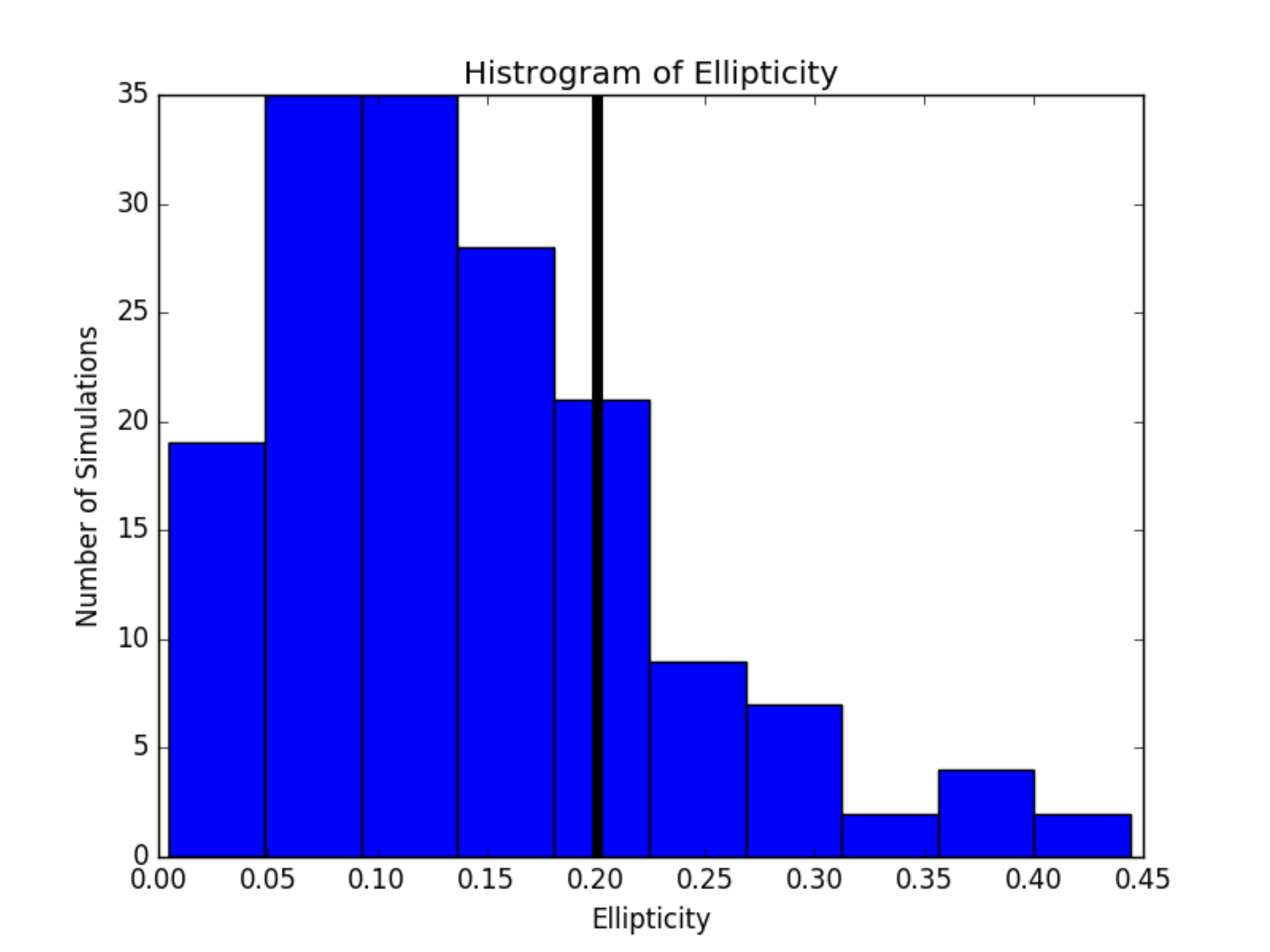}
  \caption{Histogram of the ellipticities of the final merger objects.
    The black line at $\epsilon = 0.2$ denotes the commonly adopted
    value for cE galaxies.  Most simulations exhibit objects with
    $\epsilon$ values below or equal to $0.2$, i.e. resemble the shape
    of most cE galaxies observed.}
  \label{fig:Ellipticity}
\end{figure}

The ellipticity of an object is defined as 
\begin{eqnarray}
  \label{eq:eps}
  \epsilon & = & 1 - \frac{b}{a}
\end{eqnarray}
where $a$ and $b$ are the semi-major and minor axis of an ellipse,
respectively.  The ellipticity of our objects is obtained from the
function {\sc Ellipse} in {\sc IRAF} using the 2D pixel-maps (see e.g.\
left panel of Fig.~\ref{fig:SimulationExample}) we compute.  As
ellipticities can change with radius, we use the ellipticity value of
the isophote closest to $100$~pc.

\begin{figure*}
  \centering
  \includegraphics[width=1\textwidth]{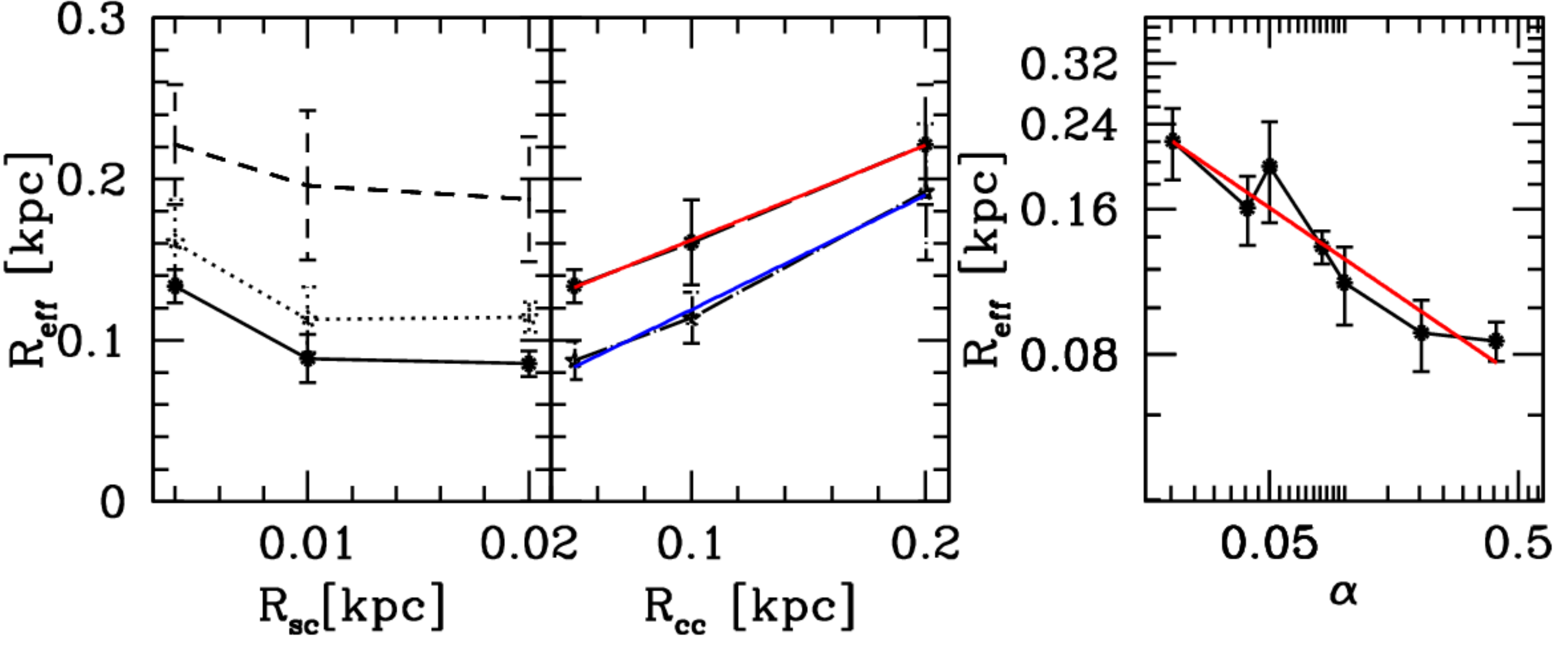}
  \caption{Effective radius $R_{\rm eff}$ of the merger objects.  In
    the left panel we plot the effective radius as function of the
    size of the constituents $R_{\rm sc}$.  Solid line connects
    simulations with $R_{\rm cc} = 50$~pc, dotted line connects
    simulations with $R_{\rm cc} = 100$~pc and dashed lines show
    $R_{\rm cc} = 200$~pc simulations.  We see that the final
    effective radius is almost independent of the choice of the size
    of the SCs.  Only for very small SCs the effective radius is
    slightly higher but only within the error-bars.  The middle panel
    shows $R_{\rm eff}$ as function of the distribution size $R_{\rm
      cc}$.  The upper curve is from $R_{\rm sc} = 4$~pc simulations
    and for the lower curve both $R_{\rm sc}$-values of $10$ and
    $20$~pc are used.  In the right panel we plot $R_{\rm eff}$
    against the filling factor parameter $\alpha$.}
  \label{fig:Reff}
\end{figure*}

In the literature it is usually considered that compact ellipticals
have no significant ellipticity \citep[e.g.][]{Chilingarian2015}.  The
ellipticity of M32 is about E2 which means that the smaller axis is
about 20\% shorter than its larger axis.  The exact ellipticity of M32
is 0.17 \citep{Kent1987}. 

The ellipticity of the final object is not related to any of our
simulation parameters ($R_{\rm gal}$, $R_{\rm sc}$ and $R_{\rm cc}$).
In Fig.~\ref{fig:Ellipticity} one can see that almost all the
simulations have ellipticities in the range of cEs,
i.e. $\epsilon = 0.2$ or less. 

\subsection{Effective radius of the final object (R$_{\rm eff}$)}
\label{sec:Reff}

To obtain the effective radius $R_{\rm eff}$ of our final objects
(after $10$~Gyr of evolution) we determine the radial surface
brightness profile in logarithmically spaced concentric rings out to a
maximum radius of $1.5$~kpc.  As our objects exhibit low ellipticities
(see Sect.~\ref{sec:shapes}), the use of rings instead of ellipses
poses only small errors.  We perform this calculation in all three 
main projections ($x-y$, $x-z$ and $y-z$) and calculate the mean
values to avoid chance alignments.  These mean values are now fitted
using a Sersic profile as described in Eq.~\ref{eq:sersic}.

In our simulations we see no dependency of $R_{\rm eff}$ with the
choice of the galactic distance of our cluster complex.  Again, this
can be explained by the fact that all our models are initially tidally 
under-filling and therefore the gravitational pull of the galactic
potential has no or only very small influence on the formation and
evolution of the merger object.  Therefore, we can neglect the
galactic distance in the following analysis and bin all simulations
with the same internal parameters independent of their distance to the
galaxy to obtain better statistics.

The same is true if we compare simulations with $N_{0} = 64$ and
$128$.  As most of the SCs/UCDs in the CC merge to build the final
object, its internal parameters are only marginally dependent on the
initial number of constituents and we are able to increase the
statistical significance of our results further by binning the
simulations independent of their initial $N_{0}$ as well.

In Fig.~\ref{fig:Reff} we show the resulting effective radius as
function of the remaining parameters $R_{\rm sc}$, $R_{\rm cc}$ and
their combination, i.e.\ the filling factor $\alpha$.  The left panel
shows $R_{\rm eff}$ as function of $R_{\rm sc}$.  We see a slight
trend that with a larger size of the constituents we obtain merger
objects which are more concentrated.  But this trend is well within
the calculated $1\sigma$ error-bars.  Especially, between $R_{\rm sc}
= 10$~pc and $20$~pc we see no significant difference any longer.
This means, as long as we build our merger object out of extended
constituents, their size does not influence the scale radius of the
final object.  Only if concentrated initial SCs are used then we may
see a slower merging process \citep[see][for a more detailed
explanation]{Fellhauer2002} and a more extended final object.

The middle panel of Fig.~\ref{fig:Reff} shows only two different lines
as we choose to bin the simulations with $R_{\rm sc} = 10$ and $20$~pc
here (top line shows $R_{\rm sc} = 4$~pc results, bottom line $R_{\rm
  sc} = 10$ and $20$~pc).  A very clear linear trend is visible that
with a larger size of the distribution, we see a larger final object.
We fit a line to the results and obtain as slope $0.59 \pm 0.01$ for
$R_{\rm sc} = 4$~pc, shown as the red line and $0.71 \pm 0.06$ for the
other results (blue line).  As we focus more on simulations with
extended initial objects we can give the following relation as a rule
by thumb:
\begin{eqnarray}
  \label{eq:reff1}
  R_{\rm eff} & \approx & 0.7 \cdot R_{\rm cc}.
\end{eqnarray}

Finally, in the right panel we show the dependence of our results on
the ratio of both input parameters: $\alpha$ (see
Eq.~\ref{eq:alpha}).  In the double-logarithmic plot the results of
our simulations follow a linear trend, i.e.\ a power law dependence.
If we fit a power-law to our results we get $-0.35 \pm 0.05$ as
power-law index, i.e.\ we can write:
\begin{eqnarray}
  \label{eq:reff2}
  R_{\rm eff} & \propto & \alpha^{-1/3}.
\end{eqnarray}

\begin{figure*}
  \centering
  \includegraphics[width=1\textwidth]{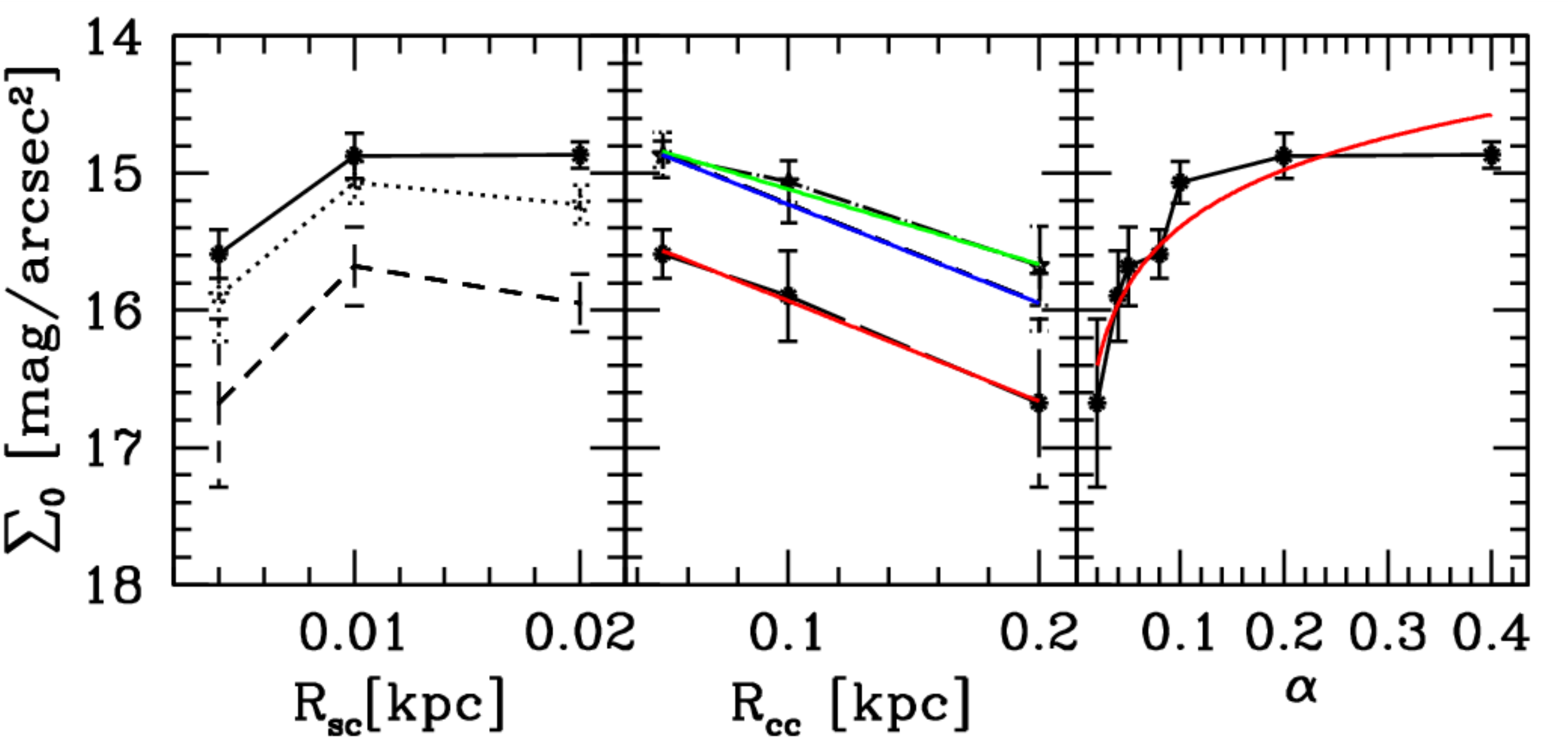}
  \caption{Central Surface Brightness $\Sigma_{0}$. In the left panel
    we show the central surface brightness as function of the size of
    the constituents $R_{\rm sc}$.  The lines are the same as in
    Fig.~\ref{fig:Reff}.  The middle panel shows $\Sigma_{0}$ as
    function of the size of the distribution $R_{\rm cc}$.  Bottom line
    (red) is for $R_{\rm sc} = 4$~pc, middle line (blue) for $R_{\rm
      sc} = 10$~pc and top line (green) for $R_{\rm sc} = 20$~pc.
    In the right panel we plot $\Sigma_{0}$ as function of the filling
    factor $\alpha$.}
  \label{fig:CSB}
\end{figure*}

This result implies that a larger 'filling factor' $\alpha$ leads to
more compact merger objects.  An explanation could be deduced when
compared to the results found in \citet{Fellhauer2002}.  There, larger
values of $\alpha$ lead to a change in the merging behaviour, i.e.\ the
SCs rather merged with the fast growing central object.  If $\alpha$
is low then the constituents rather merge with each other first and
the central merger object appeared later.  As a merger between two
objects (dry merger - without gas) always leads to a more extended
object than the two initial ones, the above mentioned change in
merging behaviour can explain that we find more compact merger objects
when starting with less compact objects to begin with.  Here the initial 
constituents form first very extended secondary objects, which then 
merge and build an even more extended central object.

Comparing the results to the quoted effective radii for compact
ellipticals in the literature, which ranges between $80$ and $200$~pc,
only the simulations with the smallest scale-length for the
constituents ($R_{\rm sc} = 4$~pc) together with the largest size of
the cluster complex distribution ($R_{\rm cc} = 200$~pc) are just
outside this window (see left panel of Fig.~\ref{fig:Reff}).  This
points into the direction that, to obtain cEs (at least in this formation
scenario), we need compact, massive CCs with extended objects to have a
fast and efficient merger process.  Therefore, we need the merging of
UCD-like objects to form a cE rather than compact SCs.

\subsection{Central Surface Brightness ($\Sigma_{0}$)}
\label{sec:CSB}

In this section we present the results of the central surface
brightness $\Sigma_{0}$ of our merger objects.  As explained before we
construct smoothed 2D pixel maps of our objects with their surface
density measured in M$_{\odot}$\,pc$^{-2}$.  These values we convert
into mag.\,arcsec$^{-2}$ using a generic mass-to-light ratio of
unity.  As magnitudes are a logarithmic scale, a factor of a few,
obtained by using your favourite pass-band and a mass-to-light ratio
for a mainly old population in that exact band, will alter our results
only marginally, in extreme cases to about one magnitude fainter.

Once the 2D map is constructed, we use the brightest, central pixel to
deduce the central surface brightness of our objects.  As our pixel
size is $20$~pc, our models have a higher resolution as most
observations of distant cEs could obtain.  This could lead to another
source of higher values of $\Sigma_{0}$ deduced from our simulations
than actually observed.

The values reported in Tab.~\ref{tab:results} with their errors are
obtained by taking a mean value from the three simulations with the
same parameters but different random realisations.

Looking at the results, no dependency on the distance $R_{\rm gal}$ to
the centre of the galaxy is visible.  The explanation is the same as
for the effective radii, our models are all tidally under-filling at
the beginning and the gravitational forces of the host galaxy are of
rather minor importance.  Furthermore, as in the previous section we
do not see any strong dependence on the number of initial objects
$N_{0}$.  We are therefore able to use all simulations independent of
$R_{\rm gal}$ and $N_{0}$ to calculate the mean values and errors,
increasing the statistical base.

In Fig.~\ref{fig:CSB} we show the central surface brightness of the
merger objects as function of the same initial parameters as in
Fig.~\ref{fig:Reff} and we see exactly the same trend as with the
effective radius.  For very small ($R_{\rm sc} = 4$~pc) constituents
we get more extended objects with lower central brightnesses as for
the more extended initial objects ($R_{\rm sc} = 10$ and $20$~pc),
which again do not differ significantly (left panel).  The dashed line
shows the results for $R_{\rm cc} = 50$~pc, the dotted line shows
$R_{\rm cc} = 100$~pc results and finally the solid line represents
$R_{\rm cc} = 200$~pc.

\begin{figure*}
  \centering
  \includegraphics[width=1\textwidth]{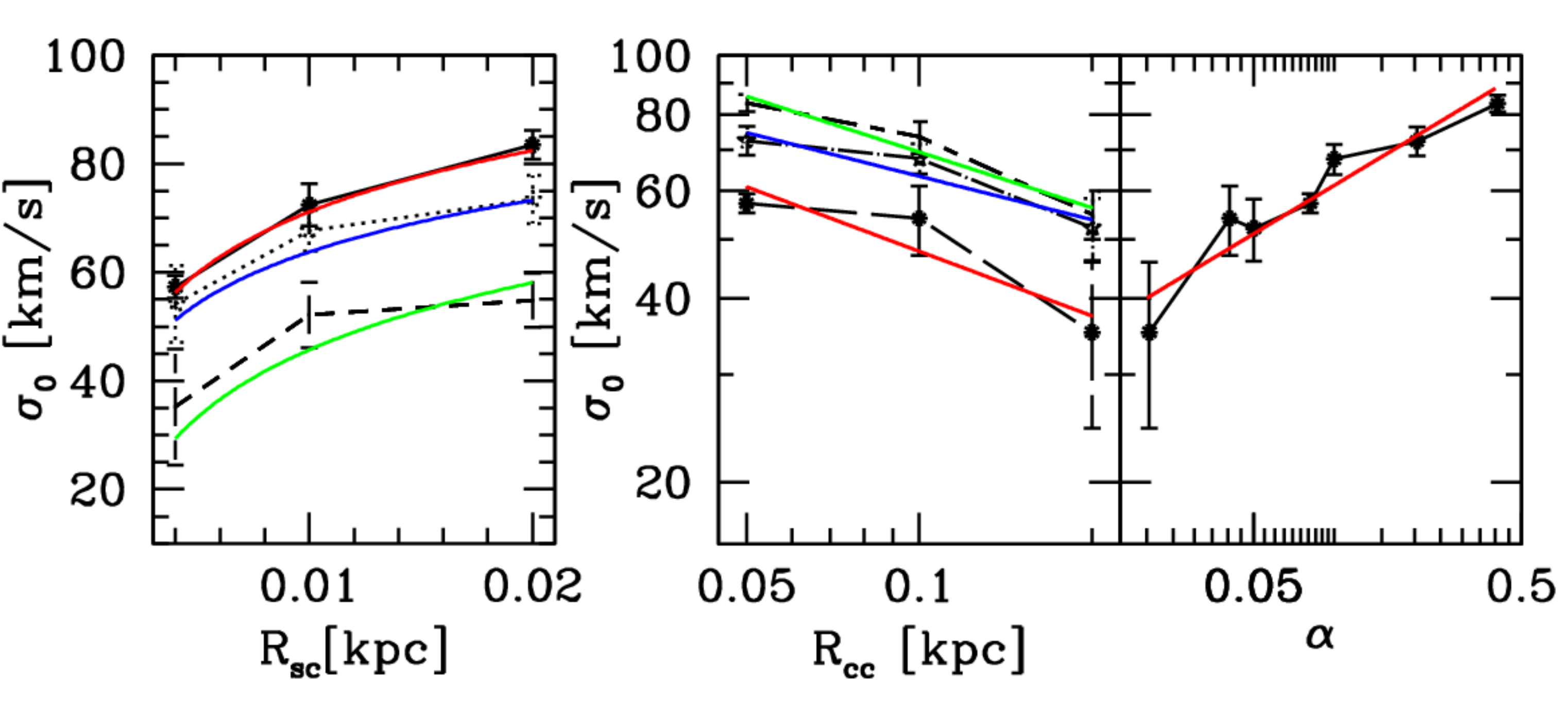}
  \caption{Central velocity dispersion $\sigma_{0}$.  Left panel:
    Central velocity dispersion as function of $R_{\rm sc}$.  Middle
    panel: $\sigma_{0}$ as function of $R_{\rm cc}$.  Right panel:
    $\sigma_{0}$ as function of $\alpha$.  Lines are the same as in
    Fig.~\ref{fig:CSB}.} 
  \label{fig:CVD}
\end{figure*}

The middle panel also shows the same trend as before: larger initial
distributions form less concentrated merger objects which in turn have
lower central brightnesses.

The central surface brightness correlates linearly with the size of
the CC.  The slopes of the fitting curves are $7.3 \pm 0.4$, $5.5 \pm
0.6$ and $7.19 \pm 0.02$ for $R_{\rm sc} = 4$ (red curve), $10$ (blue)
and $20$~pc (green) respectively ($R_{\rm cc}$ measured in kpc). 

The right panel again shows the surface brightness as function of the
parameter $\alpha$.  The trend is as expected that with larger alphas
we obtain brighter, i.e.\ more concentrated, objects.  In contrast to
$R_{\rm eff}$, which seems to decrease further with $\alpha$, here
there might be a trend to a kind of maximum central luminosity of about
$14.9$~mag\,arcsec$^{-2}$.  If there is an equivalent minimal
effective radius at higher $\alpha$ values as probed in this study
here, remains open.  This has to be investigated in a possible follow
up of this study.

Fitting a power law to the results we obtain that
\begin{eqnarray}
  \label{eq:csb1}
  \Sigma_{0} & \propto & \alpha^{-0.039 \pm 0.007}.
\end{eqnarray}

The surface brightness of known cEs ranges approximately from $14$ to
$20$~mag\,arcsec$^{-2}$ \citep[see table~1 in][]{Chilingarian2007}
\citep[table~2 in][]{Huxor2013}.  In that sense all of our simulations
match the observational constraints found for cE galaxies.

\subsection{Central Velocity Dispersion ($\sigma_{0}$)}
\label{sec:CVD}

In this section we focus on the dynamics of the merger objects.  As a
benchmark to compare our results with observations we use the central
velocity dispersion $\sigma_{0}$ of our models.  The value is obtained
summing all line-of-sight velocities of the particles located in the
densest pixel, we use to determine the central surface brightness.

Tab.~\ref{tab:results} shows that the results again do not depend on
the distance to the galaxy $R_{\rm gal}$ and the used number of SCs
$N_{0}$ and we are able to group simulations only differing by these
parameters together.

In Fig.~\ref{fig:CVD} we show the dependency of $\sigma_{0}$ on the
parameters $R_{\rm sc}$ (left panel), $R_{\rm cc}$ (middle panel) and
the combination of the two parameters $\alpha$ (right panel).  Colours
and lines are the same as in Fig.~\ref{fig:CSB}.

We see a straightforward trend, which is expected, taking the previous
results into account.  As the final mass of our cEs is approximately
the same, the velocity dispersion should depend on the size of the
object only, i.e.\ more concentrated, smaller objects should exhibit
larger velocity dispersion.  Exactly this is visible in our results.
As explained before, smaller constituents lead to a slower merging and
more extended objects and therefore exhibit lower velocity
dispersions, i.e.\ $R_{\rm sc}$ and $\sigma_{0}$ are correlated.  On
the contrary, larger distributions lead to larger objects and
henceforth to small velocity dispersions, the two quantities are
anti-correlated. 

The fitting lines shown in the figures are power-laws.  In the middle
panel we obtain the following exponents for the decrease: $-0.4 \pm
0.2$, $-0.24 \pm 0.08$ and $-0.30 \pm 0.07$ for the different choices
of $R_{\rm sc} = 20$, $10$ and $4$~pc respectively.  As we have found
a close to linear dependency of $R_{\rm eff}$ on $R_{\rm cc}$ we would
expect from simple stellar dynamical arguments (virial equilibrium)
\begin{eqnarray}
 \sigma & \sim & R_{\rm cc}^{-0.5}. \nonumber
\end{eqnarray}
We see this relation only holds if we start with large enough
constituents, i.e.\ UCD-like objects.

In the right panel we show how $\sigma_{0}$ varies with $\alpha$.  As
$\alpha$ is proportional to $R_{\rm sc}$ and anti-proportional to
$R_{\rm cc}$ we expect that $\sigma_{0}$ should have a strong
correlation with $\alpha$.  A power-law fitted to the results gives a
slope of $0.26 \pm 0.04$ and so we could approximately write
\begin{eqnarray}
  \label{eq:sigma1}
  \sigma_{0} & \propto & \alpha^{1/4}.
\end{eqnarray}

The range in observed velocity dispersions in cE galaxies is rather
large.  In their table~2 \citet{Ferre2017} report for NGC\,2970 $47.7
\pm 1.6$~km\,s$^{-1}$ and for PSC012519 a value of $222 \pm
2.3$~km\,s$^{-1}$.  The central velocity dispersion of M32 is reported
in \citet{Marel1998} to be $76 \pm 10$~km\,s$^{-1}$.

If we pose a minimum velocity dispersion of $40$~km\,s$^{-1}$ to call
a merger object a cE galaxy, we see that all our simulations except
the parameter combination of $R_{\sc} = 4$~pc and $R_{\rm cc} =
200$~pc are reproducing objects with velocity dispersions similar to
cE galaxies.  Again only the parameter set which leads to a slow
merging process is not able to reproduce the necessary values for
cEs. 

\section{Discussion and Conclusions}
\label{sec:conc}

We perform simulations in the merging star cluster scenario
using a very high total mass of $10^{9}$~M$_{\odot}$ for the cluster
complex.  The merging star cluster scenario is successful in
reproducing the formation of extended star clusters and UCDs.  With
the chosen high total mass, which is an order of magnitude higher than
used for the formation of massive UCDs we want to open up a new
formation scenario for compact ellipticals like M32.  We argue that in
the early, more gas-rich Universe we might have stronger star burst
events as seen today in interacting galaxies, which might lead to the
formation of a compact star forming region, i.e.\ a cluster complex,
producing $10^{9}$~M$_{\odot}$ in star clusters, forming a bound
entity.  

In the present Universe we do see recently formed UCD-like
  objects with luminous and dynamical masses of $10^{8}$~M$_{\odot}$.
  As an example we point to W3 in NGC~7252 \citep{Maraston2004}.  At
  an age of $300$ to $500$~Myr this UCD-like object is not formed
  through any stripping or other destructive channel.  It has the age
  of the interaction of the merger remnant of the host system and
  therefore was formed in a star-formation burst during the
  interaction.  As star-formation is believed to be higher in the
  past \citep[e.g.][]{Shibuya2016,Tacchella2018}, we think it is
  reasonable to assume that CCs with masses similar to cEs could have
  formed.

  \citet{Renaud2015} pointed out that in their simulations of
  interacting galaxies the most massive star-forming regions are found
  close to the centre of the merging pair, i.e.\ making these massive
  objects ideal candidates to sink to the centre due to dynamical
  friction, while \citet{Duc2000} find observational evidence of tidal
  dwarf galaxies at the tip of the tidal tails with masses exceeding
  $10^{9}$~M$_{\odot}$. 

  \citet{Fellhauer2005} reported about the observability of the
  merging star cluster formation scenario.  As this scenario is a
  very rapid process, the authors concluded that only if W3 in
  NGC~7252 is at its lower age limit ($300$~Myr), there might be a
  chance to observe the final stages of the ongoing merging process.
  At the higher age limit ($500$~Myr) the merging process is over and
  the object is dynamically indistinguishable from other formation
  scenarios.  Therefore, for old objects like M32 we do not expect to
  find any dynamical tracer of its formation process any longer, which
  would confirm our scenario.

  Some groups argue that the presence of tidal tails associated
  with UCDs are a clear sign of a past stripping process but
  \citet{Bruens2009} report similar tails in the merging star cluster
  scenario.  The presence or absence of a massive black hole in the
  centre could be a hint of a more massive galaxy origin.  Indeed,
  \citet[e.g.][]{Voggel2018} found massive black holes in UCDs around
  Centaurus A, making them excellent candidates for a more massive
  galaxy origin.  These high resolution observations are still ongoing
  but for every positive detection there is also a non-detection in a
  similar UCD present.  

  All the presented evidence makes it likely that objects like UCDs
  are not from a single formation channel but actually a mixed bag of
  objects. 

  As we have argued with examples of the less-massive UCDs, we think
  that the same line of reasoning can be applied to compact
  ellipticals as well.

There is one thing which our simple stellar dynamical models cannot
reproduce.  Observing M32 \citet{Monachesi2012} found two different
stellar populations:
\begin{itemize}
\item $\sim 40 \pm 17$\% of the mass in a 2-5 Gyr old, metal-rich
  population ([M/H]$ = 0.02 \pm 0.01$ dex)
\item $\sim 55 \pm 21$\% of the total mass in stars older than 5 Gyr,
  with slightly subsolar metallicities.  
\end{itemize}
Our simulations do not treat metallicities and and gas is not included
in our simulations.  We would speculate that, when forming a cluster
complex that massive, the expelled gas, seen e.g.\ as H-alpha bubbles
around the CCs in the Antennae \citep[e.g.][]{Whitmore1999}, stays
bound to the forming merger object and is able to fall back in,
forming at least a second generation of stars.

In \citet{Alarcon2018} we have shown in a different project, that our
results and conclusions do not change when adopting a star formation
history into our models.  So, we are confident that in this scenario
an included star formation history would also show no major
differences.

Every simulation performed in this study leads to a final bound object
in which the majority of star cluster have merged.  We analyse the
resulting object and are determining its shape, mass, effective
radius, central surface brightness and central velocity dispersion.
These values are compared with observational values of compact
ellipticals found in the literature.  

The main result of this study is, that indeed it is possible to obtain
a cE galaxy in the merging star cluster scenario, matching all the
above mentioned observables.

Furthermore, we established the following relations:
\begin{itemize}
\item The effective radius of the final object is inversely
  proportional to the size of the used star clusters.  This result
  seems at first counter-intuitive as dry merger processes, as
  happening here in this study, are always increasing the effective
  radius. 

  The reason for this anti-correlation is in the different merging
  behaviour we find for different choices of parameters.  Defining a
  'filling factor' $\alpha = R_{\rm sc} / R_{\rm cc}$, low values of
  $\alpha$, i.e.\ using small, compact star clusters, lead to the
  merging of pairs of star clusters throughout the cluster complex
  first, before finally a central object is built up.  This object now
  has a larger scale radius.

  In the case of large filling factors, the merging process is very
  fast and directly with the forming central object.  This results in
  more compact objects.
\item The effective radius is linearly proportional to the used size
  of the cluster complex.  As a rule-by-thumb we give $R_{\rm eff} =
  0.7 R_{\rm cc}$.
\item With our studied parameter range, the obtained effective radii
  are between $90$ and $260$~pc, which marks the range of cE
  galaxies.  
\item The range of ellipticities $\epsilon$ is between $0.05$ and
  $0.32$, with almost all simulations having ellipticities in the
  range of cEs ($\epsilon \leq 0.2$).  
\item We always obtain central surface brightnesses which are
  matching the values of cEs.
\item The merger objects exhibit velocity dispersions in the range of
  about $50$ to $80$~km\,s$^{-1}$, thereby matching the observational
  values.  We have in total three simulations in which the final
  velocity dispersion is too low to match cEs.  These simulations
  correspond to incomplete merging (many star clusters avoid the
  merging process and fly away) and have to be ruled out by showing a
  final mass which is too low anyway.  All of these three simulations
  have $R_{\rm sc} = 4$~pc and $R_{\rm cc} = 200$~pc, i.e.\ $\alpha =
  0.02$. 
\end{itemize}

These results point to the fact that in order to obtain an object
similar to a cE galaxy, it is more favourable to have a cluster
complex with a high filling factor, i.e.\ either small extension or
extended objects forming within.

The merging star cluster scenario is not the only way to produce a cE
galaxy.  Other theories for the formation of cEs include the tidal
stripping and truncation scenario \citep{Faber1973} either with an
elliptical galaxy with a dense core as origin
\citep[e.g.][]{Faber1973} or the bulge of a partially stripped disc
galaxy \citep[e.g.][]{Bekki2001}.  Or cEs could also have an intrinsic
origin and are the natural extension of the class of elliptical
galaxies to smaller sizes and lower luminosities \citep{Wirth1984,
  Kormendy2009}.

Nevertheless, our scenario opens up a new possible formation channel
to explain the existence of compact elliptical galaxies. \\

{\bf Acknowledgments:} 
FUZ, MF, AGAJ, DRMC and CAA acknowledge funding through Fondecyt Regular
No. 1180291 and BASAL Centro de Astrofisica y Tecnologias Afines (CATA)
AFB-170002.  MF acknowledges funding through the Concurso Proyectos 
Internacionales de Investigacion No. PII20150171.  AGAJ acknowledges 
financial support from Carnegie Observatories through the Carnegie-Chile 
fellowship.

\bibliographystyle{mn2e}

\begin{thebibliography}{99}

\bibitem[\protect\citeauthoryear{Alarcon Jara et al.}{2018}]{Alarcon2018} 
  Alarc\'on Jara, A.G.; Fellhauer, M.; Matus Carrillo, D.R.; Assmann,
  P.; Urrutia Zapata, F.; Hazeldine, J.; Aravena, C.A. 2018, MNRAS,
  473, 5015

\bibitem[\protect\citeauthoryear{Bastian et al.}{2005}]{Bastian2005}
  Bastian N., Gieles M., Efremov Y.N., Lamers H.J.G.L.M. 2005, A\&A,
  443, 79
 
\bibitem[\protect\citeauthoryear{Bekki et al.}{2001}]{Bekki2001}
  Bekki, K.; Couch, W.J.; Drinkwater, M.J.; Gregg, M.D. 2001, ApJ,
  557, 39

\bibitem[\protect\citeauthoryear{Binney \&
    Tremaine}{1987}]{Binney1987}
  Binney, J.; Tremaine S. 1987, 'Galactic Dynamics', Princton
  University Press

\bibitem[\protect\citeauthoryear{Brodie \& Larsen}{2002}]{Brodie2002}
  Brodie, J.P.; Larsen, S.S. 2002, AJ, 124, 1410

\bibitem[\protect\citeauthoryear{Br\"uns et al.}{2009}]{Bruens2009}
  Br\"uns, R.C.; Kroupa, P.; Fellhauer, M. 2009, ApJ, 702, 1268

\bibitem[\protect\citeauthoryear{Br\"uns et al.}{2011}]{Bruens2011}
  Br\"uns, R.C.; Kroupa, P.; Fellhauer, M.; Metz, M.; Assmann,
  P. 2011, A\&A, 529, 138

\bibitem[\protect\citeauthoryear{Br\"uns \& Kroupa}{2012}]{Bruens2012}
  Br\"uns, R.C.; Kroupa, P. 2012, A\&A, 547, 65

\bibitem[\protect\citeauthoryear{Burkert et al.}{2005}]{Burkert2005}
  Burkert, A.; Brodie, J.; Larsen, S. 2005, ApJ, 628, 231

\bibitem[\protect\citeauthoryear{Chandar et al.}{2004}]{Chandar2004}
  Chandar, R.; Whitmore, B.; Lee, M.G. 2004, ApJ, 611, 220

\bibitem[\protect\citeauthoryear{Chilingarian et
    al.}{2007}]{Chilingarian2007}
  Chilingarian, I.; Cayatte, V.; Chemin, L.; Durret, F.; Laganá,
  T. F.; Adami, C.; Slezak, E. 2007, A\&A, 466, 21

\bibitem[\protect\citeauthoryear{Chilingarian et
    al.}{2009}]{Chilingarian2009}
  Chilingarian, I.; Cayatte, V.; Revaz, Y.; Dodonov, S.; Durand, D.;
  Durret, F.; Micol, A.; Slezak, E. 2009, Science, 326, 1379

\bibitem[\protect\citeauthoryear{Chilingarian \&
    Zolotukhin}{2015}]{Chilingarian2015} 
  Chilingarian, I.; Zolotukhin, I. 2015, Science, 348, 418

\bibitem[\protect\citeauthoryear{Drinkwater et
    al.}{2000}]{Drinkwater2000} 
  Drinkwater, M.J.; Jones, J.B.; Gregg, M.D.; Phillipps, S. 2000,
  PASA, 17, 227

\bibitem[\protect\citeauthoryear{Du et al.}{2018}]{Du2018}
  Du, M.; et al.\ 2018, ApJ submitted; arXiv:1811.06778

\bibitem[\protect\citeauthoryear{Duc et al.}{2000}]{Duc2000}
  Duc, P.-A.; Brinks, E.; Springel, V.; Pichardo, B.; Weilbacher, P.;
  Mirabel, I. F. 2000, AJ, 120, 1238

\bibitem[\protect\citeauthoryear{Faber}{1973}]{Faber1973}
  Faber, S.M. 1973, ApJ, 179, 423

\bibitem[\protect\citeauthoryear{Fellhauer et
    al.}{2000}]{Fellhauer2000}
  Fellhauer, M.; Kroupa, P.; Baumgardt, H.; Bien, R.; Boily, C.M.;
  Spurzem, R.; Wassmer, N. 2000, NewA, 5, 305

\bibitem[\protect\citeauthoryear{Fellhauer et
    al.}{2002}]{Fellhauer2002}
  Fellhauer, M.; Baumgardt, H.; Kroupa, P.; Spurzem, R. 2002, CeMDA,
  82, 113

\bibitem[\protect\citeauthoryear{Fellhauer \&
    Kroupa}{2002a}]{Fellhauer2002a} 
  Fellhauer, M.; Kroupa, P. 2002a, MNRAS, 330, 642

\bibitem[\protect\citeauthoryear{Fellhauer \&
    Kroupa}{2002b}]{Fellhauer2002b} 
  Fellhauer, M.; Kroupa, P. 2002b, AJ, 124, 2006

\bibitem[\protect\citeauthoryear{Fellhauer \&
    Kroupa}{2003}]{Fellhauer2003}
  Fellhauer, M.; Kroupa, P. 2003, Ap\&SS, 284, 643

\bibitem[\protect\citeauthoryear{Fellhauer \&
    Kroupa}{2005}]{Fellhauer2005} 
  Fellhauer, M.; Kroupa, P. 2005, MNRAS, 359, 223

\bibitem[\protect\citeauthoryear{Ferr\'e-Mateu et
    al.}{2017}]{Ferre2017} 
  Ferr\'e-Mateu, A.; Forbes, D.A.; Romanowsky, A.J.; Janz, J.; Dixon,
  C. 2017, MNRAS, 473, 1819

\bibitem[\protect\citeauthoryear{Graham}{2002}]{Graham2002}
  Graham, A.W. 2002, ApJ, 568, 13

\bibitem[\protect\citeauthoryear{Hilker et al.}{1999}]{Hilker1999}
  Hilker, M.; Infante, L.; Kissler-Patig, M.; Richtler, T. 1999,
  A\&AS, 134, 75 

\bibitem[\protect\citeauthoryear{Huxor et al.}{2005}]{Huxor2005}
  Huxor, A.P.; et al. 2005, MNRAS, 360, 1007

\bibitem[\protect\citeauthoryear{Huxor et al.}{2011}]{Huxor2011}
  Huxor, A.P.; Phillipps, S.; Price, J.; Harniman, R. 2011, MNRAS,
  414, 3557

\bibitem[\protect\citeauthoryear{Huxor et al.}{2013}]{Huxor2013}
  Huxor, A.P.; Phillipps, S.; Price, J. 2013, MNRAS, 430, 1956

\bibitem[\protect\citeauthoryear{Janz et al.}{2016}]{Janz2016}
  Janz, J.; et al. 2016, MNRAS, 456, 617

\bibitem[\protect\citeauthoryear{Kent}{1987}]{Kent1987}
  Kent, S.M. 1987, AJ, 94, 306

\bibitem[\protect\citeauthoryear{Kormendy et al.}{2009}]{Kormendy2009} 
  Kormendy, J.; Fisher, D.B.; Cornell, M.E.; Bender, R. 2009, ApJS,
  182, 216

\bibitem[\protect\citeauthoryear{Kormendy \&
    Bender}{2012}]{Kormendy2012} 
  Kormendy, J.; Bender, R. 2012, ApJS, 198, 2

\bibitem[\protect\citeauthoryear{Larsen \&
    Richtler}{1999}]{Larsen1999}
  Larsen, S.S.; Richtler, T. 1999, A\&A, 345, 59

\bibitem[\protect\citeauthoryear{Maraston et al.}{2004}]{Maraston2004}
  Maraston C., Bastian N., Saglia R. P., Kissler-Patig M., Schweizer
  F., Goudfrooij P., 2004, A\&A, 416, 467

\bibitem[\protect\citeauthoryear{Messier}{1784}]{Messier1784}
  Messier, C. 1784, Connoissance des Temps ou des Mouvements
  C\'elestes, 'Catalogue des N\'ebuleuses et des Amas d'\'Etoiles',
  pp. 227-267 

\bibitem[\protect\citeauthoryear{Misgeld \&
    Hilker}{2011}]{Misgeld2011}
  Misgeld, I.; Hilker, M. 2011, MNRAS, 414, 3699

\bibitem[\protect\citeauthoryear{Mizutani et al.}{2003}]{Mizutani2003}
  Mizutani, A.; Chiba, M.; Sakamoto, T. 2003, ApJ, 589, 89

\bibitem[\protect\citeauthoryear{Monachesi et
    al.}{2012}]{Monachesi2012} 
  Monachesi, A.; et al. 2012, ApJ, 745, 97

\bibitem[\protect\citeauthoryear{Norris et al.}{2014}]{Norris2014}
  Norris, M.A.; et al. 2014, MNRAS, 443, 1151

\bibitem[\protect\citeauthoryear{Plummer}{1911}]{Plummer1911}
  Plummer, H.C. 1911, MNRAS, 71, 460

\bibitem[\protect\citeauthoryear{Renaud et al.}{2015}]{Renaud2015}
  Renaud, F.; Bournaud, F.; Duc, P.-A. 2015, MNRAS, 446, 2038

\bibitem[\protect\citeauthoryear{Sersic}{1963}]{Sersic1963}
  S\'ersic, J.L. 1963, BAAA, 6, 41

\bibitem[\protect\citeauthoryear{Shibuya et al.}{2016}]{Shibuya2016}
  Shibuya, T.; Ouchi, M.; Kubo, M.; Harikane, Y. 2016, ApJ, 821, 72

\bibitem[\protect\citeauthoryear{Tacchella et
    al.}{2018}]{Tacchella2018} 
  Tacchella, S.; Bose, S.; Conroy, C.; Eisenstein, D.J.; Johnson,
  B.D. 2018, ApJ, 868, 92
	
\bibitem[\protect\citeauthoryear{van der Marel et
    al.}{1998}]{Marel1998}
  van der Marel, R.P.; Cretton, N.; de Zeeuw, P.T.; Rix, H.-W. 1998,
  ApJ, 493, 613

\bibitem[\protect\citeauthoryear{Voggel et al.}{2018}]{Voggel2018}
  Voggel, K.T.; et al. 2018, ApJ, 858, 20

\bibitem[\protect\citeauthoryear{Whitmore et al.}{1999}]{Whitmore1999}
  Whitmore, B.C.; Zhang, Q.; Leitherer, C.; Fall, S.M. 1999, AJ, 118,
  1551 

\bibitem[\protect\citeauthoryear{Wirth \& Gallagher}{1984}]{Wirth1984}
  Wirth, A.; Gallagher, J.S., III 1984, ApJ, 282, 85

\end{thebibliography}

\label{lastpage}

\end{document}